\documentclass[apj,square,numbers]{emulateapj}
\usepackage{graphicx}
\usepackage{latexsym}
\usepackage{color}
\usepackage[colorlinks, linkcolor=blue, citecolor=blue, urlcolor=blue]{hyperref}
\usepackage{apjfonts}

\begin{document}

\title{Testing CPT Symmetry with Current and Future CMB Measurements}

\author{Si-Yu Li$^{1}$}
\author{Jun-Qing Xia$^{2}$}
\author{Mingzhe Li$^{3}$}
\author{Hong Li$^{2}$}
\author{Xinmin Zhang$^{1}$}
\affil{$^1$Theory Division, Institute of High Energy Physics, Chinese Academy of Science, P. O. Box 918-4, Beijing 100049, P. R. China}
\affil{$^2$Key Laboratory of Particle Astrophysics, Institute of High Energy Physics, Chinese Academy of Science, P. O. Box 918-3, Beijing 100049, P. R. China}
\affil{$^3$Interdisciplinary Center for Theoretical Study, University of Science and Technology of China, Hefei, Anhui 230026, P. R. China}

\begin{abstract}

In this paper we use the current and future cosmic microwave background (CMB) experiments to test the Charge-Parity-Time Reversal (CPT) symmetry. We consider a CPT-violating interaction in the photon sector $\mathcal{L}_{\rm cs}\sim p_\mu A_\nu \tilde{F}^{\mu\nu}$ which gives rise to a rotation of the polarization vectors of the propagating CMB photons. By combining current CMB polarization measurements, the nine-year WMAP, BOOMERanG 2003 and BICEP1 observations, we obtain a constraint on the isotropic rotation angle $\bar{\alpha} = -2.12 \pm 1.14$ ($1\sigma$), indicating an about $2\sigma$ significance of the CPT violation. Here, we particularly take the systematic errors of CMB measurements into account. Then, we study the effects of the anisotropies of the rotation angle [$\Delta{\alpha}({\bf \hat{n}})$] on the CMB polarization power spectra in detail. Due to the small effects, the current CMB polarization data can not constrain the related parameters very well. We obtain the 95\% C.L. upper limit of the variance of the anisotropies of the rotation angle $C^\alpha(0) < 0.035$ from all the CMB datasets. More interestingly, including the anisotropies of rotation angle could lower the best fit value of $r$ and relax the tension on the constraints of $r$ between BICEP2 and Planck. Finally, we investigate the capabilities of future Planck polarization measurements on $\bar{\alpha}$ and $\Delta{\alpha}({\bf \hat{n}})$. Benefited from the high precision of Planck data, the constraints of the rotation angle can be significantly improved.

\end{abstract}

\maketitle


\section{Introduction}

In the standard model of particle physics and some of its extensions, the Charge-Parity-Time Reversal (CPT) symmetry has a fundamental status. Probing its violation is an important way to search for the new physics beyond the standard model. Up to now, CPT symmetry has passed a number of high-precision experimental tests and no definite signal of its violation has been observed in the laboratory. So, the present CPT violating effects, if exist, should be very small to be amenable to the laboratory experimental limits.

However, the CPT symmetry could be dynamically violated in the expanding universe. This has many interesting applications. For instances, in the literatures \citep{Li:2002,Li:2003,Li:2004,Feng:2005,Li:2007,Davoudiasl:2004gf}, the cosmological CPT violation has been used to generate the matter-antimatter asymmetry in the early universe based on the mechanism proposed in \citet{Cohen}.  The salient feature of these models is that the CPT violating effects at present are too small to be detected by the laboratory experiments, but large enough in the early universe to account for matter-antimatter asymmetry. More importantly, these types of CPT violating effects could be accumulated to be observable for the cosmological probes \citep{Feng:2005,Li:2007,Feng:2006}. With the accumulation of high-quality observational data, especially those from the cosmic microwave background (CMB) experiments, cosmological observation becomes a powerful way to test the CPT symmetry.

Simply the cosmological CPT violation in the photon sector can be modeled by the coupling between photons and an external field $p_{\mu}$ through the Chern-Simons (CS) term $\mathcal{L}_{\rm CS}\sim p_{\mu}A_{\nu}\tilde{F}^{\mu\nu}$. Here $\tilde{F}^{\mu\nu}=(1/2)\epsilon^{\mu\nu\rho\sigma}F_{\rho\sigma}$ is the dual of the electromagnetic tensor. This coupling is gauge invariant if $\partial_{\nu} p_{\mu}=\partial_{\mu} p_{\nu}$. This is possible if $p_{\mu}$ is a constant field over the spacetime or arises from the derivative of a cosmic scalar field $\phi$. The scalar field $\phi$ is identified as the dark energy in the quintessential baryo-/leptogenesis \citep{Li:2002,Li:2003} and as the Ricci scalar $R$ in the gravitational baryo-/leptogenesis \citep{Li:2004,Davoudiasl:2004gf}. The CS term violates the Lorentz and CPT symmetries spontaneously at the background in which $p_{\mu}$ is nonzero. {One of its physical consequences is that the polarization vector of the photon is no longer transported parallel along the light ray:
\begin{eqnarray}
k^{\mu}\nabla_{\mu}Q+\nabla_{\mu}k^{\mu}Q&=&2p_{\mu}k^{\mu}U\nonumber\\
k^{\mu}\nabla_{\mu}U+\nabla_{\mu}k^{\mu}U&=&-2p_{\mu}k^{\mu}Q~.
\end{eqnarray}
Here, $Q$ and $U$ are Stokes parameters describing the linear polarizations of the radiation field. They are not conserved due to the CS term. The vector $k^{\mu}$ is the four-vector of the photon.}
The rotation of the polarization direction of electromagnetic waves propagating over large distances 
\begin{equation}
(Q'\pm iU')={\rm exp}(\pm i2\alpha)(Q\pm iU)~.\label{rotation}
\end{equation}
The rotation angle is the integral of $p_{\mu}$ along the photon's trajectory from the source of light ($s$) to the observing point ($o$) \citep{Li:2008}
\begin{equation}\label{ro}
\alpha=\int^s_o p_{\mu}dx^{\mu}~.
\end{equation}
The related phenomena called ``cosmological birefringence'' has the effect of changing the polarization of the radiation from radio galaxies and quasars \citep{Carroll:1990,Carroll:1998zi}. For CMB, it has the effect to covert part of E-modes polarization to B-modes polarization, and especially it has the possibility to produce TB and EB correlations even though these are absent before recombination in the traditional CMB theory. In the case of isotropic rotation angle, denoted as $\bar{\alpha}$, the full set of the rotated CMB spectra (denoted by primes) were first obtained in \citet{Feng:2006}
\begin{eqnarray}\label{rotation}
C_{\ell}^{\rm 'TB} &=& C_{\ell}^{\rm TE}\sin(2\bar{\alpha})~, \nonumber\\
C_{\ell}^{\rm 'EB} &=&
\frac{1}{2}(C_{\ell}^{\rm EE}-C_{\ell}^{\rm BB})\sin(4\bar{\alpha})~,\nonumber\\
C_{\ell}^{\rm 'TE} &=& C_{\ell}^{\rm TE}\cos(2\bar{\alpha})~,\nonumber\\
C_{\ell}^{\rm 'EE} &=& C_{\ell}^{\rm EE}\cos^2(2\bar{\alpha}) +
C_{\ell}^{\rm BB}\sin^2(2\bar{\alpha})~,\nonumber\\
C_{\ell}^{\rm 'BB} &=& C_{\ell}^{\rm BB}\cos^2(2\bar{\alpha}) +
C_{\ell}^{\rm EE}\sin^2(2\bar{\alpha})~,
\end{eqnarray}
while the CMB temperature power spectrum remains unchanged. These formulae combined with CMB data can be used to detect or constrain the rotation angle, i.e., the signal of CPT violation. For instance one may detect $\bar{\alpha}$ by searching for the distinctive TB correlation \citep{Lue:1999}. But as was first pointed out in \citet{Feng:2005} that in the CMB polarization experiments the EB spectrum will be the most sensitive for probing the signal of CPT violation. Another important feature of this model is that it provides a new mechanism to produce the CMB B-modes polarization, alternative to the primordial gravitational waves and weak lensing. This can be seen from the last equation of Eqs. (\ref{rotation}), even the primordial B-modes is absent, sizable CMB BB power spectrum can be obtained from the EE power spectrum through the rotation. Based on Eqs. (\ref{rotation}) the first evidence on the rotation angle in terms of the full CMB datasets was done in \citet{Feng:2006} and stimulated many interests in this field (see \citet{Li:2007,Xia:2008a,WMAP5,Xia:2008b,Wu:2009,Brown:2009,WMAP7,Xia:2010,Liu:2006, Xia:Planck,Xia:Sys,WMAP9,Others1,Others2,Others3,Others4,Others5,Others6,Others7,Others8}, and references within). These studies showed that current CMB experiments have the possibility to detect the rotation angle at the level of $\mathcal{O}(1^\circ)$, provide a powerful method to test the fundamental Lorentz and CPT symmetries.

Recently, the {\it Background Imaging of Cosmic Extragalactic Polarization} (BICEP1) \citep{bicep2013} collaboration has released their high precision three-year data of the CMB temperature and polarization including the TB and EB power spectra. Another CMB experiment, the {\it Q/U Imaging ExperimenT} (QUIET) \citep{quiet}, also published the CMB polarization power spectra at 95 GHz with the EB power spectrum. Furthermore, the nine-year WMAP (WMAP9) \citep{WMAP9}, BOOMERanG 2003 (B03) \citep{B03,B031,B032} and QUaD \citep{Quad} also provided the TB and EB polarization power spectra. Thus, it is important and necessary to combine these new data together to detect or constrain the rotation angle and to test the CPT symmetry.

However, as first pointed out in \citet{Li:2008} the rotation angle is generally direction dependent or anisotropic, denoted by [$\alpha(\hat{\bf n})$]. For instance if the external field $p_{\mu}\propto \partial_{\mu}\phi$, arising from a cosmic scalar field $\phi$, the rotation angle is determined by the distribution of this field on the last scattering surface. Usually this distribution is not homogeneous because $\phi$ as a dynamical field must fluctuate around its uniform background. Hence the CMB photons coming from different directions would undergo different rotations. As first studied in \citet{Li:2008}
the anisotropies of the rotation angle will introduce corrections or distortions to the spectra (\ref{rotation}) from isotropic rotation. At a later time, the authors of \citet{Kamionkowski:2008} and \citet{Gluscevic:2009} also studied the direction dependence of rotation angle and developed a different formalism to measure the anisotropies of the rotation angle and constructed the minimum-variance estimator (similar method can also be found in \citet{Yadav:2009eb}). They considered the non-gaussian signal and the correlation between different $\ell$ and $m$ of the rotated polarization angular momentum introduced by the rotation. Then they applied the estimator to constraint the anisotropic rotation angle and found no evidence of non-zero power spectrum of rotation angle within $3\sigma$ \citep{Gluscevic:2012}. Recently, \citet{Li:2013} performed a non-perturbative calculation of the rotated power spectra and made constraints on the anisotropic rotation angle and the shape of its power spectrum in terms of the CMB data. According to these results, there was no significant evidence for a nonzero rotation angle up to now.

Following the previous works, in this paper we will revisit this problem by attaching more importance on the effects of direction-dependent rotations on the CMB power spectra. We will perform a global analysis on them using the latest CMB polarization data, as well as the future simulated CMB data. The structure of the paper is as follows: in section \ref{data} we describe the current and future simulated datasets we use. Section \ref{result} contains our main results from the current observations and future measurements, while section \ref{summary} is dedicated to the conclusions and discussions.


\section{CMB Datasets}\label{data}

\subsection{Current Datasets}

In our calculations we mainly use the full data of WMAP9 temperature and polarization power spectra \citep{WMAP7}. The WMAP9 polarization data are composed of TE/TB/EE/BB/EB power spectra on large scales ($2\leq \ell\leq23$) and TE/TB power spectra on small scales ($24\leq \ell \leq800$), while the WMAP9 temperature data are only used to set the underlying cosmology. For the systematic error, the WMAP instrument can measure the polarization angle to within $\pm1.5$ deg of the design orientation \citep{WMAP7Page03,WMAP7Page07}. In the computation we use the routines for computing the likelihood supplied by the WMAP team. Besides the WMAP9 information, we also use some small-scale CMB observations.

The {\it BOOMERanG dated January 2003 Antarctic flight} \citep{B03} measures the small-scale CMB polarization power spectra in the range of $150\leq \ell\leq1000$. Recently, the BOOMERanG collaboration re-analyzed the CMB power spectra and took into account the effect of systematic errors rotating the polarization angle by $-0.9\pm0.7$ deg \citep{Pagano:2009}.

The BICEP1 \citep{bicep2013} and {\it QU Extragalactic Survey Telescope at DASI} (QUaD) \citep{Quad} collaborations released their high precision data of the CMB temperature and polarization including the TB and EB power spectra. These two experiments, locating at the South Pole, are the bolometric polarimeters designed to capture the CMB information at two different frequency bands of $100$GHz and $150$GHz, and on small scales -- the released three-year BICEP1 data are in the range of $21\leq \ell\leq335$ \citep{bicep2013}; whereas the QUaD team measures the polarization spectra at $164\leq \ell\leq 2026$, based on an analysis of the observation in the second and third season \citep{Wu:2009,Brown:2009}. They also provide the systematic errors of measuring the polarization angle, $\pm1.3$ deg and $\pm0.5$ deg, for BICEP1 and QUaD observations, respectively.

Very recently, the BICEP2 collaboration announced the detection of CMB B-modes polarization and released the data in the $150$GHz frequency band \citep{bicep2014}. However, they claimed that the EB power spectra are only used for the self-calibration of the detector polarization orientations \citep{keating}. Any polarization rotation has been removed from the results. Therefore, in our calculations we only use the BICEP2 data to constrain the anisotropies of the rotation angle.

Finally, we have the CMB polarization power spectra at 95 GHz from the second season QUIET observation. Using two pipelines to analyze the data, they characterized the EB power spectrum between $\ell =$ 25 and 975 and gave the total systematic error in the EB power spectrum \citep{quiet}.

\subsection{Future Datasets}

The Planck collaboration has released the first cosmological papers providing the high resolution, full sky, CMB maps. Due to the improved precision, this new Planck data have constrained several cosmological parameters at the few percent level. However, this Planck data do not include the CMB polarization information and the rotation angle has nothing to do with the CMB temperature power spectrum. Therefore, current CMB measurements are still not accurate enough to verify the possible CPT violation. In order to improve the constraints on the rotation angle, we follow the method given in \citet{Xia:2008a,Xia:Planck} and simulate the CMB polarization power spectra with the assumed experimental specifications of the Planck \citep{Planck:total} polarization measurement. We choose the best-fit model from the Planck data \citep{Planck:fit} as the fiducial model.

\begin{table}[t]
\caption{Assumed experimental specifications. We use the CMB power spectra only at $\ell\leq2500$. The noise parameters $\Delta_T$ and $\Delta_P$ are given in units of $\mu$K-arcmin.}\label{futureCMB}
\begin{center}
\begin{tabular}{lcccccc}
\hline \hline

~Experiment~ & ~$f_{\rm sky}$~ & ~$\ell_{\rm max}$~ & ~(GHz)~ &
~$\theta_{\rm FWHM}$~ & ~$\Delta_T$~ & ~$\Delta_P$~ \\

\hline

~PLANCK & 0.80 & 2500 & 100 & 9.5' &  6.8 & 10.9 \\
        &      &      & 143 & 7.1' &  6.0 & 11.4 \\
        &      &      & 217 & 5.0' & 13.1 & 26.7 \\

\hline \hline
\end{tabular}
\end{center}
\end{table}

In Table \ref{futureCMB} we list the assumed experimental specifications of the future Planck polarization measurement. The likelihood function is $\mathcal{L}\propto \exp(-\chi_{\rm eff}^2/2)$ and
\begin{equation}\label{simu}
\chi^2_{\rm eff}=\sum_{\ell}(2\ell+1)f_{\rm
sky}\left(\frac{A}{|\bar{C}|}+\ln\frac{|\bar{C}|}{|\hat{C}|}+3\right)~,
\end{equation}
where $f_{\rm sky}$ denotes the observed fraction of the sky in the real experiments, $A$ is defined as:
\begin{eqnarray}\label{AAA}
A &=&
\hat{C}^{TT}_{\ell}(\bar{C}^{EE}_{\ell}\bar{C}^{BB}_{\ell}-(\bar{C}^{EB}_{\ell})^2)+\hat{C}^{TE}_{\ell}(\bar{C}^{TB}_{\ell}\bar{C}^{EB}_{\ell}-\bar{C}^{TE}_{\ell}\bar{C}^{BB}_{\ell})\nonumber\\
  &+& \hat{C}^{TB}_{\ell}(\bar{C}^{TE}_{\ell}\bar{C}^{EB}_{\ell}-\bar{C}^{TB}_{\ell}\bar{C}^{EE}_{\ell})+\hat{C}^{TE}_{\ell}(\bar{C}^{TB}_{\ell}\bar{C}^{EB}_{\ell}-\bar{C}^{TE}_{\ell}\bar{C}^{BB}_{\ell})\nonumber\\
  &+& \hat{C}^{EE}_{\ell}(\bar{C}^{TT}_{\ell}\bar{C}^{BB}_{\ell}-(\bar{C}^{TB}_{\ell})^2)+\hat{C}^{EB}_{\ell}(\bar{C}^{TE}_{\ell}\bar{C}^{TB}_{\ell}-\bar{C}^{TT}_{\ell}\bar{C}^{EB}_{\ell}) \nonumber\\
  &+& \hat{C}^{TB}_{\ell}(\bar{C}^{TE}_{\ell}\bar{C}^{EB}_{\ell}-\bar{C}^{EE}_{\ell}\bar{C}^{TB}_{\ell})+\hat{C}^{EB}_{\ell}(\bar{C}^{TE}_{\ell}\bar{C}^{TB}_{\ell}-\bar{C}^{TT}_{\ell}\bar{C}^{EB}_{\ell})\nonumber\\
  &+& \hat{C}^{BB}_{\ell}(\bar{C}^{TT}_{\ell}\bar{C}^{EE}_{\ell}-(\bar{C}^{TE}_{\ell})^2)~,
\end{eqnarray}
and $|\bar{C}|$ and $|\hat{C}|$ denote the determinants of the theoretical and observed data covariance matrices respectively,
\begin{eqnarray}\label{CCC}
|\bar{C}|&=&\bar{C}^{TT}_{\ell}\bar{C}^{EE}_{\ell}\bar{C}^{BB}_{\ell}+2\bar{C}^{TE}_{\ell}\bar{C}^{TB}_{\ell}\bar{C}^{EB}_{\ell}
           -\bar{C}^{TT}_{\ell}(\bar{C}^{EB}_{\ell})^2\nonumber\\
         & &-\bar{C}^{EE}_{\ell}(\bar{C}^{TB}_{\ell})^2-\bar{C}^{BB}_{\ell}(\bar{C}^{TE}_{\ell})^2~,\nonumber\\
|\hat{C}|&=&\hat{C}^{TT}_{\ell}\hat{C}^{EE}_{\ell}\hat{C}^{BB}_{\ell}+2\hat{C}^{TE}_{\ell}\hat{C}^{TB}_{\ell}\hat{C}^{EB}_{\ell}
           -\hat{C}^{TT}_{\ell}(\hat{C}^{EB}_{\ell})^2\nonumber\\
         & &-\hat{C}^{EE}_{\ell}(\hat{C}^{TB}_{\ell})^2-\hat{C}^{BB}_{\ell}(\hat{C}^{TE}_{\ell})^2~.
\end{eqnarray}
The likelihood has been normalized with respect to the maximum likelihood $\chi^2_{\rm eff}=0$, where $\bar{C}^{\rm XY}_{\ell}=\hat{C}^{\rm XY}_{\ell}$.


\section{Numerical Results}\label{result}

In our study we make a global analysis {to all the power spectra of the CMB data we have mentioned above} with the public available Markov Chain Monte Carlo package {\tt CosmoMC} \citep{Lewis:2002}, which has been modified to compute the non-zero TB and EB power spectra discussed above. We assume the purely adiabatic initial conditions and impose the flatness condition motivated by inflation. Our basic parameter space is: ${\bf P} \equiv (\omega_{b}, \omega_{c}, \Omega_\Lambda, \tau, n_{s}, A_{s}, r)$, where $\omega_{b}\equiv\Omega_{b}h^{2}$ and $\omega_{c}\equiv\Omega_{c}h^{2}$ are the physical baryon and cold dark matter densities relative to the critical density, $\Omega_\Lambda$ is the dark energy density relative to the critical density, $\tau$ is the optical depth to re-ionization, $A_{s}$ and $n_{s}$ characterize the primordial scalar power spectrum, $r$ is the tensor to scalar ratio of the primordial spectrum. For the pivot of the primordial spectrum we set $k_{\rm s0}=0.002\,$Mpc$^{-1}$. Furthermore, in our analysis we include the CMB lensing effect, which also produces B-modes from E-modes \citep{lensing}, when we calculate the theoretical CMB power spectra.

\subsection{Isotropic Rotation}

Firstly, we consider the constraint on the direction independent rotation angle $\bar{\alpha}$, induced by the CS term, from the current CMB measurements. As we know, this rotation angle is accumulated along the journey of CMB photons, and the constraints on the rotation angle depends on the multipoles $\ell$ \citep{Liu:2006}. \citet{WMAP5,WMAP7} found that the rotation angle is mainly constrained from the high-$\ell$ polarization data, and the polarization data at low multipoles do not affect the result significantly. Therefore, in our analysis, we assume a constant rotation angle $\bar{\alpha}$ at all multipoles. Further, we also impose a conservative flat prior on $\bar{\alpha}$ as, $-\pi/2\leq\bar{\alpha}\leq\pi/2$.

\begin{figure}[t]
\begin{center}
\includegraphics[scale=0.3]{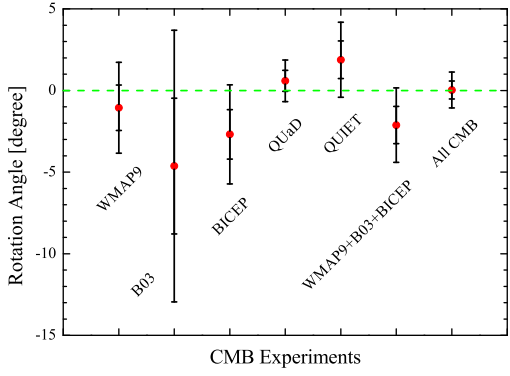}
\caption{Median values (red points) and $1\sigma$ and $2\sigma$ limits on the isotropic rotation angle $\bar{\alpha}$ obtained from different data combinations. The horizonal dashed line denotes the case $\bar{\alpha}=0$. \label{figure:alpha}}
\end{center}
\end{figure}

Following previous works \citep{Pagano:2009,Xia:Sys}, in this paper we consider the possible systematic errors of CMB measurements by including two rotation angles, $\bar{\alpha}$ and $\beta$, in order to take into account the real rotation signal and the systematic errors. Therefore, in the analyses of this subsection we have five free parameters: the direction independent rotation angle $\bar{\alpha}$ and four systematic errors, $\beta_{\rm WMAP9},\beta_{\rm B03},\beta_{\rm BICEP1},\beta_{\rm QUaD}$, for four CMB observations, respectively. And we impose priors on these four systematic errors:
\begin{eqnarray}
\beta_{\rm WMAP9}=0.0 \pm 1.5~{\rm deg}&,&~~\beta_{\rm
B03}=-0.9 \pm 0.7~{\rm deg}~,\nonumber\\
\beta_{\rm BICEP1}~=0.0 \pm 1.3~{\rm deg}&,&~~\beta_{\rm QUaD}=0.0 \pm
0.5~{\rm deg}~,
\end{eqnarray}
and marginalize over them to constrain the rotation angle. For the QUIET experiment, since they already provided the systematic errors of EB power spectrum, we directly include this information into our analysis \citep{quiet}. In Figure \ref{figure:alpha} we present current constraints on $\bar{\alpha}$ from the WMAP9, B03, BICEP1, QUaD and QUIET CMB polarization power spectra with CMB systematic errors.

Using the latest WMAP9 power spectra data at all multipoles $\ell$ and the prior of $\beta_{\rm WMAP9}$, we obtain the constraint on the rotation angle: $\bar{\alpha}=-1.06\pm1.39$ deg at $68\%$ confidence level, which is quite consistent with that obtained from the WMAP team \citep{WMAP9} and is a significant improvement over the WMAP3 \citep{Xia:2008a} and WMAP5 \citep{WMAP5,Xia:2008b} results. Similarly, we obtain the constraint on $\bar{\alpha}$ from the B03 polarization data: $\bar{\alpha}=-4.63\pm4.16$ deg, with the CMB systematic effect included. The results are in good agreement with the previous results \citep{Feng:2006,Pagano:2009}.

In our previous work \citep{Xia:2010}, we reported that the two-year BICEP1 data \citep{bicep} favored a non-zero $\bar{\alpha}$ at about $2.4\sigma$ confidence level, due to the clear bump structure in the BICEP1 TB and EB power spectra at $\ell \sim 150$. Even including the impact of systematic effect, the significance is still larger than $2\sigma$ from the BICEP1 data alone \citep{Xia:Sys}. Recently, the BICEP1 collaboration released the three-year data and paid particular attention on the constraint of the rotation angle \citep{bicep2013}. They carefully discussed the effects of systematic errors, the Polarization Angle Calibration and Differential Beam Effects, on the constraint of the rotation angle and obtained the $\pm1.3$ deg systematic uncertainty on the orientation calibration. Therefore, we revisit the limit on $\bar{\alpha}$ from the new BICEP1 data and obtain the constraint at $68\%$ C.L. is:
\begin{eqnarray}
\bar{\alpha}=-2.69\pm1.52~{\rm deg}~(68\%~{\rm C.L.})~.
\end{eqnarray}
The significance of non-zero rotation angle reduces to $1.8\sigma$, due to the large systematic uncertainty of BICEP1 data. When WMAP9 and B03 data are added to the BICEP1 data, the constraint on $\bar{\alpha}$ gets tightened:
\begin{eqnarray}
\bar{\alpha}=-2.12\pm1.14~{\rm deg}&~&(68\%~{\rm C.L.})~,\nonumber\\
 -4.30 < \bar{\alpha} < 0.15~{\rm deg}&~&(95\%~{\rm C.L.})~,\label{threeconst}
\end{eqnarray}
which implies $\bar{\alpha}\neq0$ at about $2\sigma$ confidence level, when considering systematic effects of these three CMB measurements.

We also constrain the rotation angle from the QUaD and QUIET polarization data. Similarly with previous results, we use the QUaD and QUIET data and obtain the constraint at 68\% confidence level: $\bar{\alpha}=0.59\pm0.64$ deg and $\bar{\alpha}=1.88\pm1.15$ deg, respectively. When comparing with the result of WMAP9+B03+BICEP1 [Eq.(\ref{threeconst})], there is still a $\sim2\,\sigma$ tension, which needs to be taken care of in the further investigation. By combining all these CMB polarization data together and including their systematic effects, we obtain the tightest constraint: $\bar{\alpha}=0.03\pm0.55$ deg ($68\%$ C.L.).

Finally, we use the new BICEP2 polarization data to constrain the isotropic rotation angle and obtain the constraint: $\bar{\alpha}=0.12\pm0.16$ deg (68\% C.L.), due to the self-calibration of the detector polarization orientations \citep{Li2014}. This constraint is clearly expected and consistent with the use of the self-calibration method on the BICEP2 data.

\subsection{Anisotropies of Rotation Angle}

In this subsection, we briefly review the basics of the effects of the direction dependent rotation angle on the CMB power spectra \citep{Li:2013} and perform the global analyses on the related parameters from the current CMB measurements.

Firstly, we decompose the CMB temperature and polarization fields in terms of the spin-weighted spherical harmonics \citep{Seljak:1996}:
\begin{eqnarray}
T(\hat{\bf n})&=&\sum_{\ell m}T_{\ell m}Y_{\ell m}(\hat{\bf n})\nonumber\\
(Q\pm iU)(\hat{\bf n})&=&\sum_{\ell m} (E_{\ell m}\pm iB_{\ell m})~_{\pm2}Y_{\ell m}(\hat{\bf n})~.
\end{eqnarray}
Then the two parity eigenstates $E_{\ell m}$ and $B_{\ell m}$ have the parities $(-1)^\ell$ and $(-1)^{\ell+1}$ respectively. In traditional CMB theory, the TB and EB cross-correlations vanish.
As usual in the linear perturbation theory the rotation angle is decomposed into the isotropic part and the fluctuations
$\alpha(\hat{\bf n})\equiv\bar{\alpha}+\Delta\alpha(\hat{\bf n})$.
The anisotropies $\Delta\alpha(\hat{\bf n})$ considered as a random field can be decomposed in terms of (scalar) spherical harmonics on the full sky:
\begin{equation}
\Delta\alpha(\hat{\bf n})=\sum_{\ell m}b_{\ell m}Y_{\ell m}(\hat{\bf n})~,
\end{equation}
and it is possible to define the angular power spectrum of rotation angle under the assumption of statistical isotropy of $b_{\ell m}$
\begin{equation}
\langle b_{\ell m}b^{\ast}_{\ell'm'}\rangle=C_\ell^{\alpha\alpha}\delta_{\ell\ell'}\delta_{mm'}~.
\end{equation}
With the angular power spectrum one can calculate the two-point correlation function using the following relation
\begin{equation}
C^{\alpha}(\beta)\equiv\langle\Delta\alpha(\hat{\bf n})\Delta\alpha(\hat{\bf n'})\rangle=\sum_{\ell}\frac{2\ell+1}{4\pi}C_\ell^{\alpha\alpha} P_\ell(\cos\beta)~,
\end{equation}
where $\beta$ is the angle between these two directions, $\cos(\beta)=\hat{\bf n}\cdot\hat{\bf n}'$.

{Using the non-perturbative method, we calculate the the rotated power spectrum $C'_{\ell}$ and express them in terms of the unrotated ones $C_{\ell}$, via the computations of the rotated correlation functions \citep{Li:2013}.}
\begin{widetext}
\begin{eqnarray}\label{newCMB}
&&{C'_{\ell}}^{EE}+{C'_{\ell}}^{BB}=\exp{[-4C^{\alpha}(0)]}\sum_{\ell'}\frac{2\ell'+1}{2}(C_{\ell'}^{EE}+C_{\ell'}^{BB})
\int^{1}_{-1}d^{\ell'}_{22}(\beta)d^{\ell}_{22}(\beta)e^{4C^{\alpha}(\beta)}d\cos(\beta)\nonumber\\
&&{C'_{\ell}}^{EE}-{C'_{\ell}}^{BB}=\cos(4\bar{\alpha})\exp{[-4C^{\alpha}(0)]}\sum_{\ell'}\frac{2\ell'+1}{2}(C_{\ell'}^{EE}
-C_{\ell'}^{BB})\int^{1}_{-1}d^{\ell'}_{-22}(\beta)d^{\ell}_{-22}(\beta)e^{-4C^{\alpha}(\beta)}d\cos(\beta)\nonumber\\
&&{C'_{\ell}}^{EB}=\sin(4\bar{\alpha})\exp{[-4C^{\alpha}(0)]}\sum_{\ell'}\frac{2\ell'+1}{4}(C_{\ell'}^{EE}-C_{\ell'}^{BB})
\int^{1}_{-1}d^{\ell'}_{-22}(\beta)d^{\ell}_{-22}(\beta)e^{-4C^{\alpha}(\beta)}d\cos(\beta)\nonumber\\
&&{C'_{\ell}}^{TE}=\cos(2\bar{\alpha})\exp{[-2C^{\alpha}(0)]}\sum_{\ell'}\frac{2\ell'+1}{2}C_{\ell'}^{TE}\int^{1}_{-1}
d^{\ell'}_{02}(\beta)d^{\ell}_{20}(\beta)d\cos(\beta)\nonumber=C_{\ell}^{TE}\cos(2\bar{\alpha})e^{-2C^{\alpha}(0)}\nonumber\\
&&{C'_{\ell}}^{TB}=\sin(2\bar{\alpha})\exp{[-2C^{\alpha}(0)]}\sum_{\ell'}\frac{2\ell'+1}{2}C_{\ell'}^{TE}\int^{1}_{-1}
d^{\ell'}_{02}(\beta)d^{\ell}_{20}(\beta)d\cos(\beta)=C_{\ell}^{TE}\sin(2\bar{\alpha})e^{-2C^{\alpha}(0)}~.
\end{eqnarray}
\end{widetext}
{Here, $C^\alpha(0)\equiv\sum_\ell(2\ell+1)C_\ell^{\alpha\alpha}/4\pi$ is the variance of the anisotropies of the rotation angle. More detailed calculations are laid out in \citet{Li:2013}.}
In these calculations, we assume that the unrotated CMB field and the direction dependent rotation angle are Gaussian random fields and have the isotropic statistics. Consequently, the rotated polarizations of CMB are also statistically isotropic and there spectra have no off-diagonal terms.  We have done the average on the rotation angle field ensembles. These are different from the formalism developed in \citet{Kamionkowski:2008}. They considered a fixed rotation angle field which could break the statistical isotropy of the rotated CMB field. Therefore, some correlation functions between different $\ell$ and $m$ will be non-zero and the whole calculations become complicated. We leave them in the future work \citep{Li:future}.

\begin{figure}[t]
\begin{center}
\includegraphics[scale=0.45]{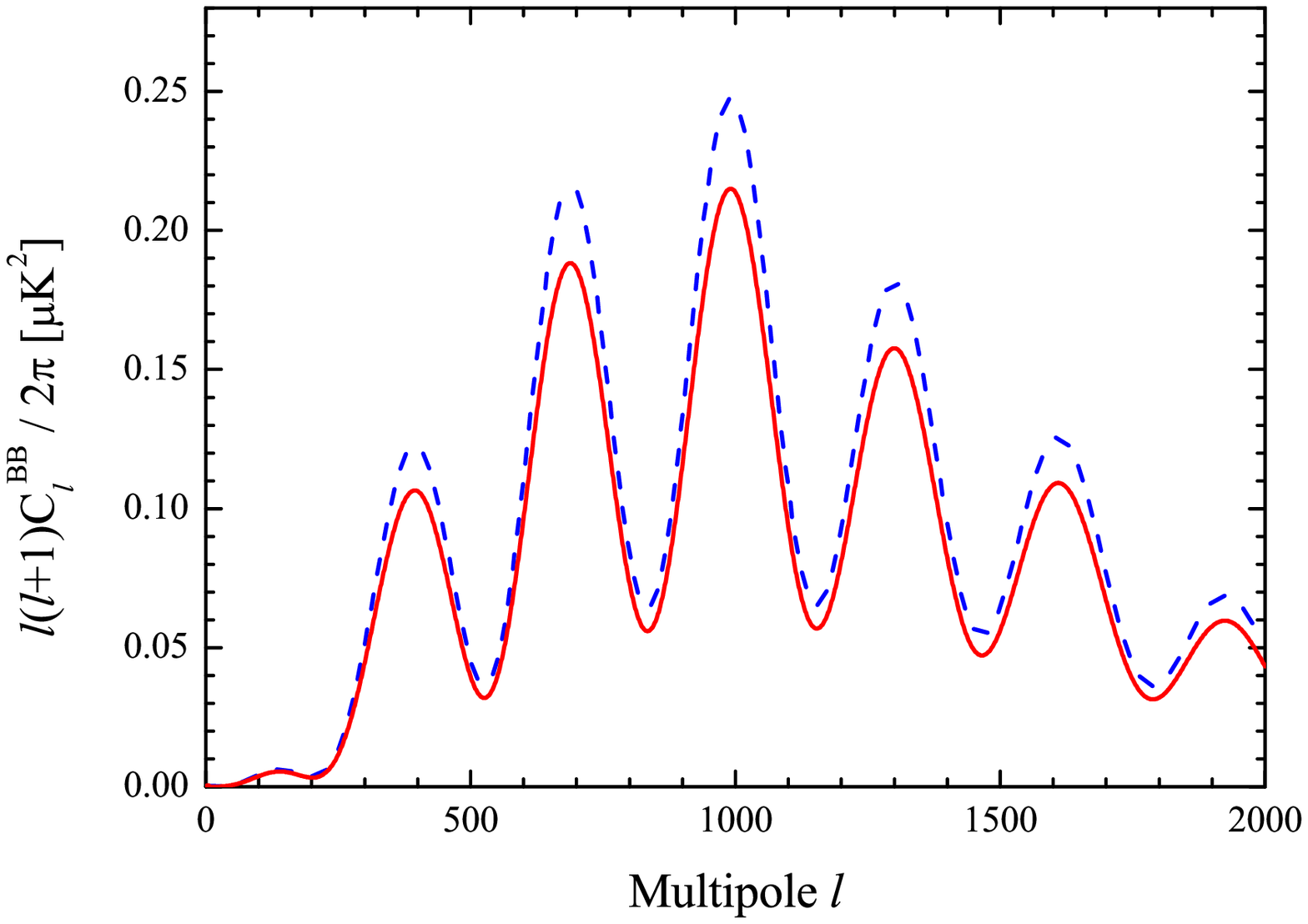}
\includegraphics[scale=0.45]{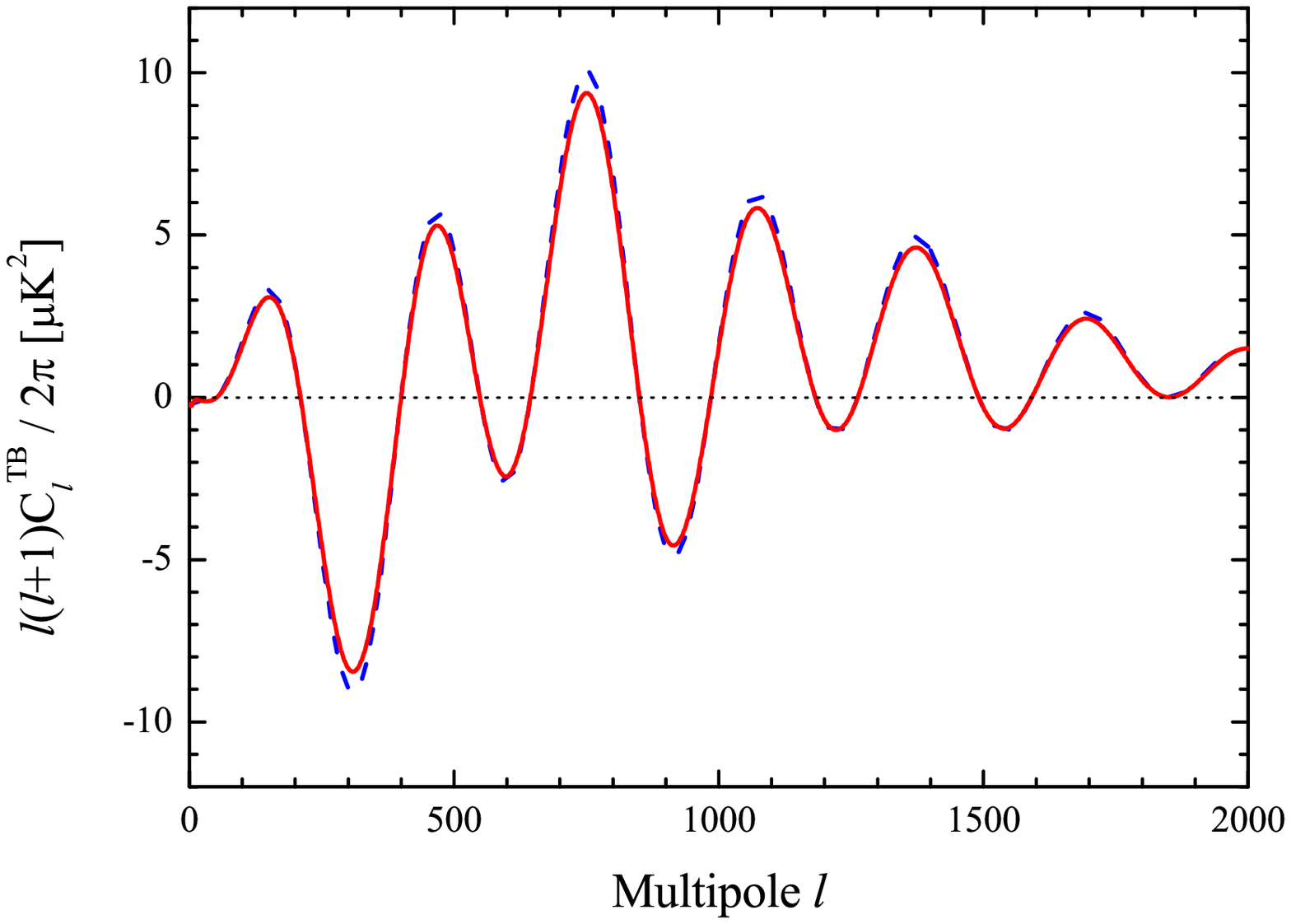}
\includegraphics[scale=0.45]{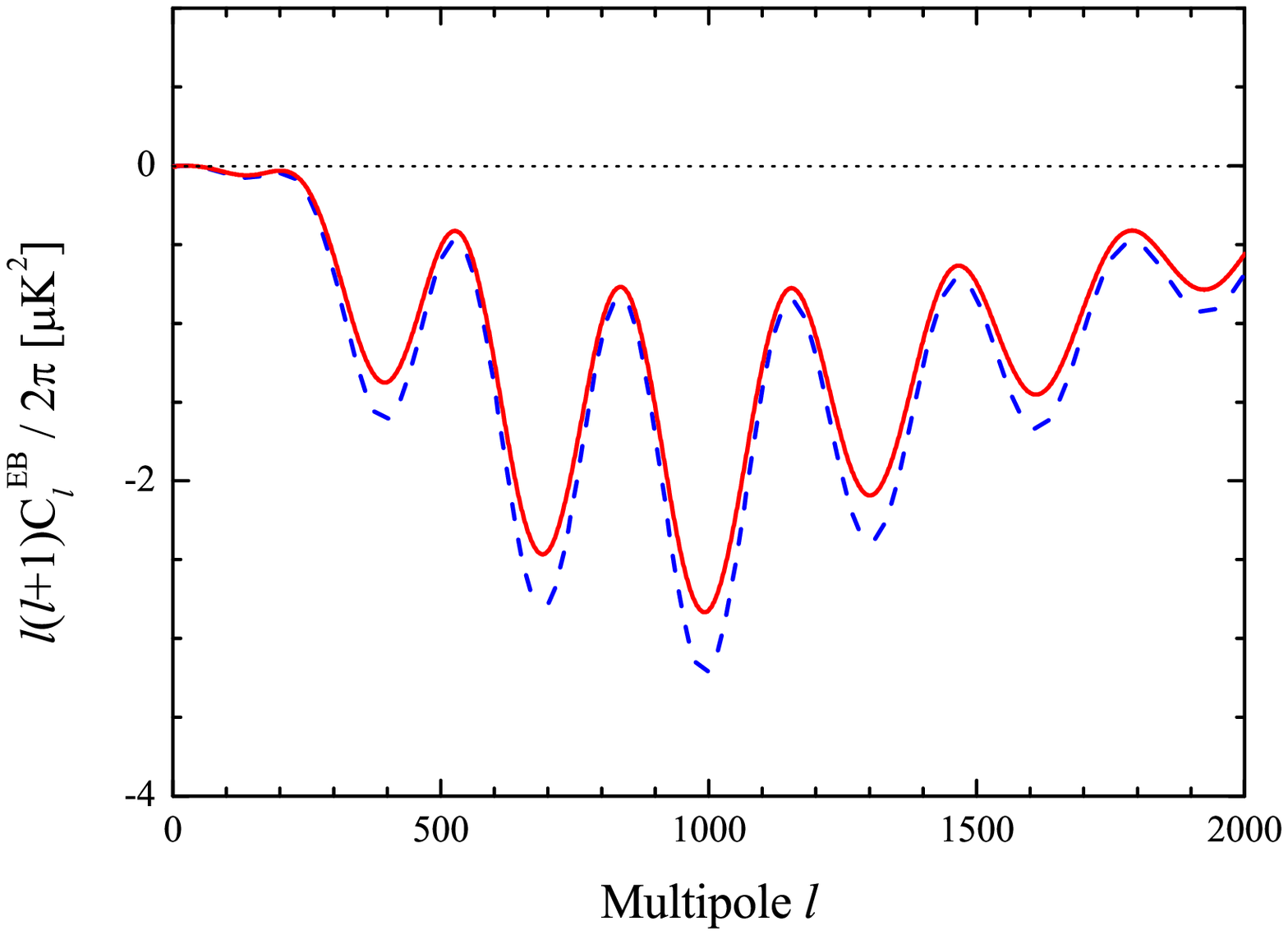}
\caption{The theoretical BB, TB and EB power spectra of the non-zero rotation angle. The red solid lines and the blue dashed lines are obtained from the $\bar{\alpha}=-2.16$ deg model and the model with $\bar{\alpha}=-2.16$ deg and $C^\alpha(0)=0.035$, respectively. The horizonal dotted lines denote the standard CMB case $\alpha\equiv0$. \label{figure:space_effect}}
\end{center}
\end{figure}

Based on equations above, we show the distortions to the CMB BB, TB and EB power spectra from the non-zero direction dependent rotation angle in Figure \ref{figure:space_effect}. Due to the non-zero variance $C^\alpha(0)$, the amplitude of BB, TB and EB power spectra are suppressed slightly when including the direction dependence of rotation angle, which means this fluctuation can be safely treated as a small effect if the isotropic rotation angle is sizeable.

\begin{figure}[t]
\begin{center}
\includegraphics[scale=0.4]{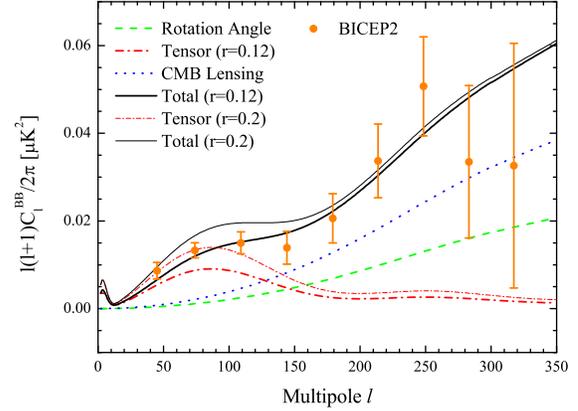}
\caption{Theoretical CMB power spectra (thick lines) for the best fit model obtained from the WMAP9+B03+BICEP2 data combination. The red thin line denotes the contribution of tensor perturbations with $r=0.2$, and the black thin line is the total CMB BB power spectrum correspondingly. For comparison, we also show the BICEP2 observational data. \label{figure:bestfit}}
\end{center}
\end{figure}

In this case, we need seven more parameters to describe the effect of the direction dependent rotation angle on CMB polarization power spectra: the variance of the two point correlation function $C^{\alpha}(0)$, and the six parameters for the binned power spectrum of rotation angle: $C_\ell^{\alpha\alpha}(i)$ ($i\in[1,6]$), which are the average values of $C_\ell^{\alpha\alpha}$ in the multipole regions $[2, 100]$, $[101, 200]$, $[201, 300]$, $[301, 400]$, $[401, 500]$, $[501, 2500]$, respectively. {We use the binned power spectrum $C_\ell^{\alpha\alpha}$ to calculate the CMB polarization power spectra to fit the CMB datasets, based on Eqs.(\ref{newCMB}).} In the calculations, we only impose physical priors on these parameters: $C^{\alpha}(0)>0$ and $C_\ell^{\alpha\alpha}(i)>0$.

\begin{figure*}[t]
\begin{center}
\includegraphics[scale=0.6]{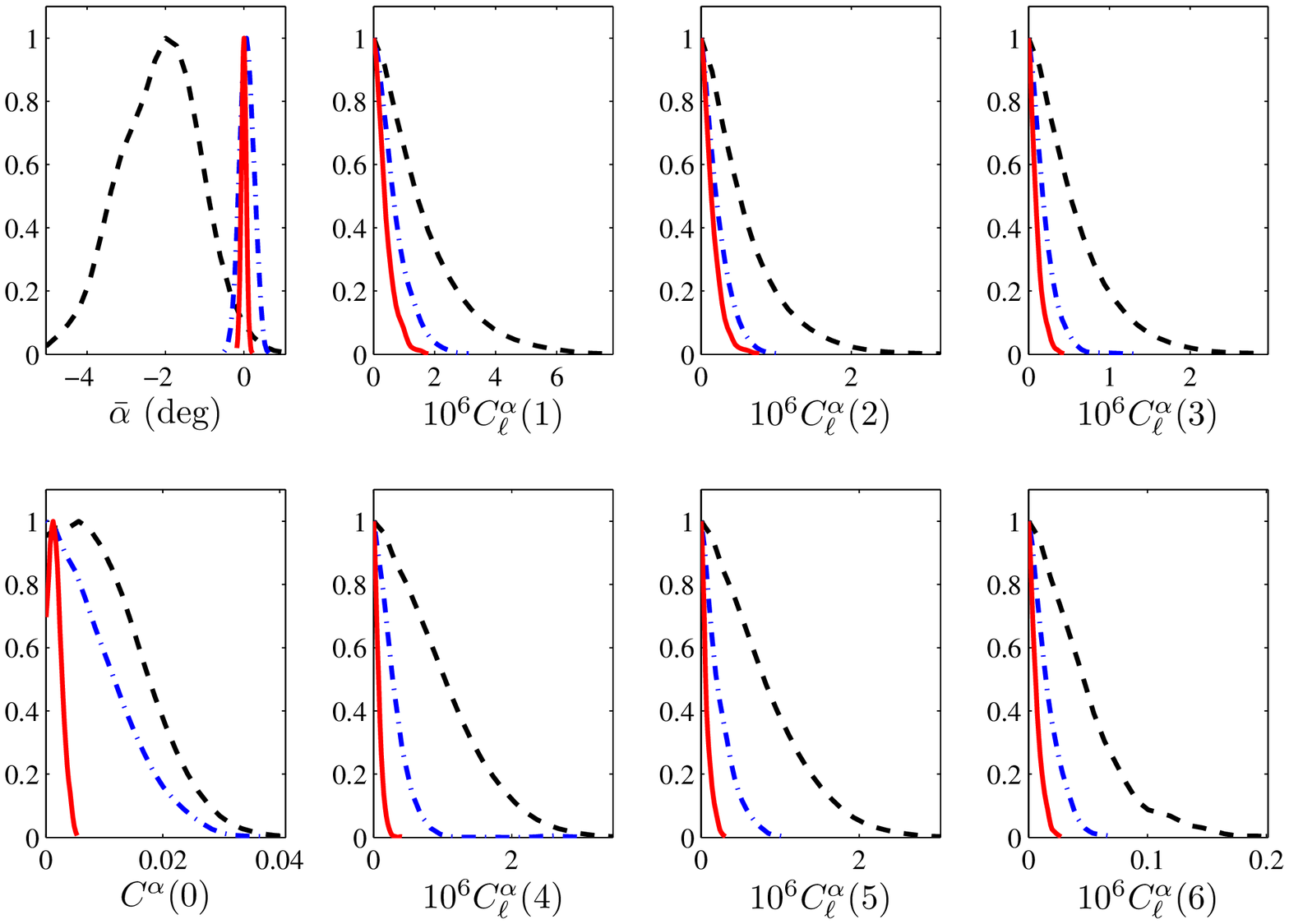}
\caption{One-dimensional distributions of parameters related to the direction dependent rotation angle ${\alpha}({\bf \hat{n}})$ from WMAP9+B03+BICEP1 (black dashed lines) and WMAP9+B03+BICEP2 (blue dash-dotted lines) data combinations and the future Planck data (red solid lines), respectively. \label{figure:planck}}
\end{center}
\end{figure*}

We use the WMAP9, B03 and BICEP1 data combination to constrain the rotation angle and obtain the limits, shown as the black dashed lines in Figure \ref{figure:planck}:
\begin{eqnarray}\label{bicep1result}
\bar{\alpha} = -2.16 \pm 1.15~{\rm deg}~&(&68\%~{\rm C.L.})~,\nonumber\\
C^\alpha(0) < 0.035~~~~~~~~ &(&95\%~{\rm C.L.})~,\nonumber\\
C_\ell^{\alpha\alpha}(1) < 3.3\times10^{-6}~&(&95\%~{\rm C.L.})~,\nonumber\\
C_\ell^{\alpha\alpha}(2) < 1.3\times10^{-6}~&(&95\%~{\rm C.L.})~,\nonumber\\
C_\ell^{\alpha\alpha}(3) < 1.1\times10^{-6}~&(&95\%~{\rm C.L.})~,\nonumber\\
C_\ell^{\alpha\alpha}(4) < 1.3\times10^{-6}~&(&95\%~{\rm C.L.})~,\nonumber\\
C_\ell^{\alpha\alpha}(5) < 1.2\times10^{-6}~&(&95\%~{\rm C.L.})~,\nonumber\\
C_\ell^{\alpha\alpha}(6) < 1.1\times10^{-7}~&(&95\%~{\rm C.L.})~.
\end{eqnarray}
We find that including the direction dependence of rotation angle does not affect the constraint on the $\bar{\alpha}$ significantly. The best fit value of $\bar{\alpha}$ is slightly smaller, since the non-zero $C^\alpha(0)$ suppresses the rotated CMB power spectra and partly cancels the effect from the non-zero $\bar{\alpha}$. This result also proves that this direction dependence of rotation angle is a small effect, comparing with the effect from the obvious non-zero $\bar{\alpha}$. The constraints on $C^{\alpha}(0)$ and $C_\ell^{\alpha\alpha}(i)$ are consistent with previous works \citep{Li:2013}.

We also include the QUaD and QUIET polarization power spectra into the analysis. Due to the high precision of the high-$\ell$ data from QUaD experiment, the constraint on the last bin of rotation angle power spectrum becomes tighter: $C_\ell^{\alpha\alpha}(6) < 6.6\times10^{-8}~(95\%~{\rm C.L.})$. We do not find the signal of the direction dependent rotation angle from the current CMB polarization power spectra.

Finally, we replace the BICEP1 data as the new BICEP2 data in the calculation. When combining with the WMAP9 and B03 polarization data, the 95\% upper limits of the parameters of direction dependence of rotation angle are shrunk by a factor of $\sim2$ (blue dash-dotted lines in Figure \ref{figure:planck}), due to the high precision of the new BICEP2 data:
\begin{eqnarray}
C^\alpha(0) < 0.023~~~~~~~~ &(&95\%~{\rm C.L.})~,\nonumber\\
C_\ell^{\alpha\alpha}(1) < 1.5\times10^{-6}~&(&95\%~{\rm C.L.})~,\nonumber\\
C_\ell^{\alpha\alpha}(2) < 0.6\times10^{-6}~&(&95\%~{\rm C.L.})~,\nonumber\\
C_\ell^{\alpha\alpha}(3) < 0.5\times10^{-6}~&(&95\%~{\rm C.L.})~,\nonumber\\
C_\ell^{\alpha\alpha}(4) < 0.6\times10^{-6}~&(&95\%~{\rm C.L.})~,\nonumber\\
C_\ell^{\alpha\alpha}(5) < 0.5\times10^{-6}~&(&95\%~{\rm C.L.})~,\nonumber\\
C_\ell^{\alpha\alpha}(6) < 0.5\times10^{-7}~&(&95\%~{\rm C.L.})~.
\end{eqnarray}
Furthermore, we obtain the constraint on the tensor-to-scalar ratio: $r=0.12\pm0.04$ (68\% C.L.), which implies the non-zero detection of the primordial CMB B-modes power spectrum. More interestingly, the median value of $r$ here is slightly lower than that reported by the BICEP2 collaboration $r=0.20$ \citep{bicep2014}, since the direction dependence of rotation angle would also contribute to the CMB BB power spectrum \citep{Li:2008, ZhaoLi2014} and partly explain the CMB B-modes data of BICEP2 \citep{Lee2014}. As we know, the Planck data could also be used to study the very early Universe and give the tight constraint on the tensor-to-scalar ratio: $r<0.11$ (95\% C.L.) \citep{Planck:fit}. Therefore, considering the direction dependence of rotation angle could lower the best fit value of $r$ and let it more consistent with the constraint from Planck data. Consequently, the tension on the constraints of $r$ between Planck and BICEP2 data has been relaxed. {Note that, in this case the constraint on the isotropic rotation angle becomes $\bar{\alpha} = -0.34 \pm 1.10~{\rm deg}~(68\%~{\rm C.L.})$. The best fit value of $\bar{\alpha}$ is quite close to zero, which means that the isotropic rotation angle can not contribute extra information on CMB BB power spectrum. Therefore, the most contribution on the explaining the discrepancy between BICEP2 and Planck's constraints comes from the anisotropic rotation angle. In Figure \ref{figure:bestfit} we show the CMB BB power spectrum of the best fit model obtained from the WMAP9+B03+BICEP2 data combination. The effect of nonzero anisotropic rotation angle does give contribution on the BB power spectrum to suppress the value of $r$. If increasing the contribution of tensor perturbations from $r=0.12$ to $r=0.2$, the theoretical prediction of CMB BB power spectrum (black thin line) is obviously higher than the BICEP2 data at $80<\ell<150$.

\subsection{Future Planck Constraints}

Since the current CMB polarization measurements can not determine the rotation angle, especially its direction dependence, conclusively, it is worthwhile discussing whether future CMB polarization data could give more stringent constraints on the rotation angle. Therefore, we simulate the future CMB power spectra with Planck to constrain the rotation angle. The fiducial model we choose is the best-fit Planck model \citep{Planck:fit}: $\Omega_{b}h^2=0.022161$, $\Omega_{c}h^2=0.11889$, $\Omega_\Lambda=0.6914$, $\tau=0.0952$, $n_{s}=0.9611$, $\log{[10^{10}A_{s}]}=3.0973$ at $k_{\rm s0}=0.05\,$Mpc$^{-1}$, and $r=0$. Here, we neglect the systematic error of future CMB measurement and the CMB lensing effect.

For the direction independent rotation angle $\bar{\alpha}$, the future Planck polarization measurement could shrink the standard deviation of rotation angle by a factor of 10, namely $\sigma(\bar{\alpha})\simeq0.06$ deg, which is consistent with our previous results \citep{Xia:2008a,Xia:Planck}. On the other hand, the future Planck mock data can also give tighter constraints on the parameters related to $\Delta{\alpha}({\bf \hat{n}})$. The 95\% C.L. upper limit of the variance becomes: $C^\alpha(0)<0.0036$, which improves the constraint by a factor of 10. In Figure \ref{figure:planck} we show the constraints on these parameters from the future Planck data (red solid lines). The high precision Planck polarization data improve the constraints on the binned rotation angle power spectrum significantly:
\begin{eqnarray}
C_\ell^{\alpha\alpha}(1) < 9.2\times10^{-7}~(95\%~{\rm C.L.})~,\nonumber\\
C_\ell^{\alpha\alpha}(2) < 3.7\times10^{-7}~(95\%~{\rm C.L.})~,\nonumber\\
C_\ell^{\alpha\alpha}(3) < 2.5\times10^{-7}~(95\%~{\rm C.L.})~,\nonumber\\
C_\ell^{\alpha\alpha}(4) < 1.9\times10^{-7}~(95\%~{\rm C.L.})~,\nonumber\\
C_\ell^{\alpha\alpha}(5) < 1.8\times10^{-7}~(95\%~{\rm C.L.})~,\nonumber\\
C_\ell^{\alpha\alpha}(6) < 1.4\times10^{-8}~(95\%~{\rm C.L.})~.
\end{eqnarray}
The future Planck data could verify the non-zero rotation angle and its direction dependence, as well as the possible cosmological CPT violation.


\section{Conclusions and Discussions}\label{summary}

Probing the signals of fundamental symmetry breakings is an important way to search for the new physics beyond the standard model. Now detecting the rotation of the CMB polarization induced by the Chern-Simons coupling is
considered as an effective and important method to test Lorentz and CPT symmetries in the physics and cosmology communities. In this paper we present constraints on the rotation angle and its anisotropies using the latest CMB polarization data, as well as the future simulated Planck data.

Following the previous works, we include the systematic effects of CMB polarization data in the analysis. Due to the larger systematic error of new BICEP1 three-year data, the significance of non-zero rotation angle reduces to around $2\sigma$, namely $\bar{\alpha}=-2.12\pm1.14~({\rm deg})$ from WMAP9+B03+BICEP1 data combination. We still find a $\sim2\,\sigma$ tension between QUaD and WMAP9+B03+BICEP1 observations. When combining all CMB polarization data together, we obtain the tightest constraint on the rotation angle at 68\% confidence level: $\bar{\alpha}=0.03\pm0.55$ deg. Furthermore, we investigated the impact of the direction dependence of the rotation angle on the CMB polarization power spectra in detail and perform a global analysis to constrain the related parameters using {\tt CosmoMC}. We found that the anisotropies of the rotation angle are just weak disturbances, namely the variance $C^\alpha(0)<0.035$ at 95\% confidence level. Due to the small effects, the current CMB polarization data can not constrain these parameters very well. The obtained results are consistent with zero.

We also consider the new BICEP2 polarization data. Since the BICEP2 collaboration uses the ``self-calibration'' for the detector polarization orientations, any polarization rotation has been removed from their data. Therefore, we mainly use this data to study the anisotropies of the rotation angle. When combining with WMAP9 and B03 data, we obtain tighter constraints, which however are still consistent with zero, on the parameters of the anisotropies of the rotation angle. Interestingly, since the direction dependence of rotation angle would also contribute the CMB BB power spectrum, considering this direction dependence could lower the best fit value of $r$ and relax the tension on the constraints of $r$ between from BICEP2 and from Planck data.

Since the current constraints on the rotation angle are not conclusive, we simulate the future Planck polarization data. We find that the future CMB data could significantly improve the constraint of $\bar{\alpha}$ by a factor of 10, as well as the parameters related to the direction dependence of the rotation angle. The future Planck data could constrain the non-zero rotation angle and its anisotropies and test the CPT symmetry more stringently.


\section*{Acknowledgements}

We acknowledge the use of the Legacy Archive for Microwave Background Data Analysis (LAMBDA). Support for LAMBDA is provided by the NASA Office of Space Science. J.-Q. X. is supported by the National Youth Thousand Talents Program and the National Science Foundation of China under Grant No. 11422323. M. L. is supported by Program for New Century Excellent Talents in University and by NSFC under Grants No. 11075074. H. L. is supported in part by NSFC under Grant Nos. 11033005 and 11322325, by the 973 program under Grant No. 2010CB83300. S. L. and X. Z. are supported in part by NSFC under Grants No. 11121092, No.11375202 and No. 11033005. The research is also supported by the Strategic Priority Research Program ``The Emergence of Cosmological Structures'' of the Chinese Academy of Sciences, Grant No. XDB09000000.

\bibliographystyle{plainnat}

\end{document}